\documentclass[12pt]{article} \textheight=23 true cm \textwidth=16 true cm
\usepackage{graphicx}
\usepackage{amsmath}
\begin{document}

\begin{center}
{\Large\bf Accelerating Universe from an Evolving  $\Lambda$ in
Higher Dimension
}\\[15 mm]
D. Panigrahi\footnote{Relativity and Cosmology Research Centre,
Jadavpur University, Kolkata - 700032, India , e-mail:
dibyendupanigrahi@yahoo.co.in , Permanent Address :
 Kandi Raj College, Kandi, Murshidabad 742137, India}
  and S. Chatterjee\footnote{Relativity and Cosmology Research Centre, Jadavpur University,
Kolkata - 700032, India, and also at NSOU, New Alipore College, Kolkata  700053,
 e-mail : chat\_ sujit1@yahoo.com\\Correspondence to : S. Chatterjee} \\[10mm]

\end{center}

\begin{abstract}
We find exact solutions in five dimensional inhomogeneous matter
dominated model with a varying cosmological constant. Adjusting
arbitrary constants of integration one can also achieve
acceleration in our model. Aside from an initial singularity our
spacetime is regular everywhere including the centre of the
inhomogeneous distribution. We also  study the analogous
homogeneous universe in (4+d) dimensions. Here an initially
decelerating model is found to give late acceleration in
conformity with the current observational demands. We also find
that both anisotropy and number of dimensions have a role to play
in determining the time of flip, in fact the flip is delayed in
multidimensional models.  Some astrophysical parameters like the
age, luminosity distance etc are also calculated and the influence
of extra dimensions is briefly discussed. Interestingly our model
yields a larger age of the universe compared to many other
quintessential models.
\end{abstract}
   ~~~~~~~KEYWORDS : cosmology; higher dimensions; varying $\Lambda$

~~~PACS :   04.20, 04.50 +h
\bigskip
\section*{1. Introduction}
   ~~~~  Recent observational evidences suggest that the present day
 universe has the critical energy density containing presumably
 $70 \%$ dark energy and about $30 \%$ dark matter, where the term
 \emph{dark} indicates a sort of invisibility. While the sceptics
 will always question the wisdom to explain data based on
 something we can not see the avalanche of data emanating from
 type Ia supernovae measurements \cite{tur}, CMB anisotropies
  \cite{per}, galactic rotation curves and surveys of galaxies,
 clusters and superclusters make the presence of dark matter and
 dark energy increasingly convincing. This situation inevitably
 forces us to pose the question : why are dark matter and dark
 energy so dark ?

 As a natural corollary to the question one is reminded of another
  `dark' stuff in physics, extra dimensions. Like dark matter and dark
  energy to explain the current quintessential behaviour the
  existence of extra dimensions is absolutely necessary in any
  attempt to unify gravity with other forces of nature beyond the
  standard model of particle physics and also to explain the vexed
  hierarchial problem of quantum mechanics in varied brane
  inspired models \cite{ran}. The recent spurt in activities in
  extra dimensions also stems from the Space-Time-Matter (STM)
  theory \cite{wes} proposed recently by Wesson and his
  collaborators \cite{scab}. As our space time is manifestly four
  dimensional in nature the extra dimensions should be `hidden' (
  or `dark' ).

  In the back drop of the above discussions it is more than
  apparent that this common feature of `invisible existence' of
  dark matter, dark energy and extra dimension points to the
  conjecture that there must be some deep underlying relationship
  and inter play among them  \cite{gu}.

   Motivated by this consideration we have in the past worked out
problems  \cite{scab} where starting from a perfect fluid in
higher dimensions obeying reasonable energy conditions
accelerating model is made possible as a consequence of the
existence of higher dimensions. Interestingly we do not have to
invoke any adhoc quintessential type of fluid by hand to achieve
the acceleration.

The motivation for the present work is somewhat different. In an
earlier communication  \cite{scbb} we have analysed an
inhomogeneous 5D spacetime with a constant $\Lambda$, and showed
that unlike the finding of Tosa  \cite{tos} the spacetime did show
the desirable feature of dimensional reduction as the usual 3D
space expands. Although in the light of observational evidences
that our universe is currently accelerating, the idea of $\Lambda$
is now an integral ingredient  in cosmological models its
introduction, nevertheless, invites serious conceptual problems.
The upper limit of $\Lambda$ from observations is about 120 orders
of magnitude below the value for the vacuum energy density
predicted from quantum field theory.

To circumvent this difficulty and a host others an essentially
phenomenological approach is generally taken where it is argued
that due to the coupling of the dynamical degrees of freedom with
matter fields of the universe, $\Lambda$ relaxes to its current
small value through the expansion of the Universe and creation of
photons \cite {vk}. From this point of view $\Lambda$ is small
because the Universe is old.

 Another salient feature of our model is its inhomogeneity. While
 the higher dimensional generalisation of the FRW models has been
 adequately addressed in the literature scant attention has been
 paid so far to address the inhomogeneous situations and the
 issues coming out from it. Moreover the work of Mustapha etal
  \cite{mus} and others indicate that there is no
 unquestionable observational evidences for spatial homogeneity.
  So investigations in inhomogeneous models is always a welcome
 step.

 Although the cosmological concordance $\Lambda CDM ~~models$ fit
  most of the observations well, as mentioned earlier, it is beset
 with theoretical difficulties. In fact whenever there is any ambiguity
 interpreting any observational results as in 1990s regarding the age
 of the universe \cite{russ} the inhomogeneous models are seriously
 considered. Similarly in the backdrop of the current accelerating
 phase of the universe the role of inhomogeneities \cite{kari} like that of extra
  dimensions as a possible ingredient of the cause of the apparent
  acceleration is being increasingly debated  \cite{rasa}. In fact averaging
   out the spatial degrees of freedom ( i.e. inhomogeneities) leads to averaged,
    effective Einstein equations (somewhat similar to what we find in higher dimensional
    models also), which contain some additional terms relating to the inhomogeneous
     distributions generally termed \emph{back reaction} in this context.
    It is argued that this so called back reaction can account for the current
   acceleration of the universe \cite{kai}. Our scope in the present work
   is very limited. While we have attempted to address the implications of
   extra dimensions in the context of current accelerating expansion in some detail
    the role inhomogeneities in our present model has been deferred to a future
work in progress.

 The paper is organised as follows: in section 2 we find an
 exact solution for a 5D inhomogeneous distribution with a
 variable $\Lambda$. Interestingly we here get
 acceleration, although at the cost of sacrificing dimensional reduction. In
 section 3 we focus on a homogeneous (d+4)-dimensional spacetime to show
 that accelerating model is possible in this case also. In section
 4 some important astrophysical parameters are discussed for our
 model to investigate how inclusion of extra dimensions influences
 the situation. Section 5 compares the $\Lambda$ varying case
 with our earlier $\Lambda$- constant situation. Here an initially
 decelerating universe ends up as an accelerating one. Here we find
 two interesting results- both anisotropy and number of dimensions
 have marked effects in determining the instant of \emph{flip}. The
 paper ends with a brief discussion in section 6.

\section*{2. The Field Equations and its integrals}

    ~~~~~~We consider a metric of (3 + d + 1)-dimensional spacetime where
both the ordinary 3-space and the extra ( or hidden) space are
inhomogeneous, isotropic and flat :
\begin{eqnarray}
  ds^{2} &=&
  dt^{2}-a^{2}\left[dr^{2} + r^{2}\left(d\theta^{2}+
  r^{2}sin^{2}\theta d\phi^{2}\right)\right] - b^{2}dy^{2}
\end{eqnarray}
$dy^{2} = \sum^{d}_{i=1}dy^{2}_{i}$ and we assume a single scale
factor for the internal dimensions, here a(r,t) and b(r,t) are
scale factors. Assuming that the matter content in this higher
dimensional space is taken to be a perfect fluid and is augmented
by the inclusion of a time varying cosmological constant we can
write down the Einstein's equations, which govern the evolution of
the ordinary 3-space as well as the extra space as

\begin{equation}
 G_{01}= 2\frac{\dot{a'}}{a}-2\frac{\dot{a}a'}{a^{2}} +  d\left(\frac{\dot{b'}}{b}
  - \frac{\dot{a} b'}{a b}\right) = 0 \\
\end{equation}
\begin{eqnarray}
 G^{1}_{1}= 2\frac{\ddot{a}}{a} + \frac{\dot{a^{2}}}{a^{2}} +
  2d \frac{\dot{a}\dot{b}}{ab} +
  \frac{d(d-1)}{2}\frac{\dot{b^{2}}}{b^{2}}+
  d\frac{\ddot{b}}{b}~~~~~~~~~~~~~~~~~~~~~~~~~~~~~~~~~~~~~~~~~~~~~~~~~~~~~~~~~~~~~~~~~~~~~\nonumber\\
~~~~~  - \frac{1}{a^{2}}\left[2\frac{a'}{ar} + 2d\frac{b'}{br} +
2d\frac{a'b'}{ab}
  +\frac{a'^{2}}{a^{2}} + \frac{d(d-1)}{2}\frac{b'^{2}}{b^{2}}\right]
  = - p + \Lambda~~~~~~~~~~~~~~~~~~~~~~~~\\
 G^{2}_{2}=G^{3}_{3}= 2\frac{\ddot{a}}{a} + \frac{\dot{a^{2}}}{a^{2}} +
  2d \frac{\dot{a}\dot{b}}{ab} +
  \frac{d(d-1)}{2}\frac{\dot{b^{2}}}{b^{2}}+ d\frac{\ddot{b}}{b} ~~~~~~~~~~~~~~~~~~~~~~~~~~~~~~
  ~~~~~~~~~~~~~~~~~~~~~~~~~~~~~~~\nonumber\\
  -\frac{1}{a^{2}}\left[\frac{a''}{a} - \frac{a'^{2}}{a^{2}} +
  \frac{a'}{ar} + d \frac{b'}{br} + \frac{d(d-1)}{2} \frac{b'^{2}}{b^{2}}
   + d \frac{b''}{b}\right] = -p + \Lambda~~~~~~~~~~~~~~~~~~~~~~~~~~\\
   G^{4}_{4}=3\frac{\ddot{a}}{a} + 3\frac{\dot{a^{2}}}{a^{2}} + 3(d-1)
   \frac{\dot{a}\dot{b}}{ab}
   + \frac{(d-1)(d-2)}{2}\frac{\dot{b^{2}}}{b^{2}} +
   (d-1)\frac{\ddot{b}}{b}~~~~~~~~~~~~~~~~~~~~~~~~~~~~~~~~~~~~~~~~~~~~~~~\nonumber\\
    -\frac{1}{a^{2}}\left[2\frac{a''}{a} -4\frac{a'}{ar}\frac{a'^{2}}{a^{2}} + (d-1)\frac{a'b'}{ab}
     + 2(d-1)\frac{b'}{br} + (d-1)(d-2)\frac{b'^{2}}{b^{2}}
     + (d-1) \frac{b''}{b}\right]~~~~~~~~~~~~~~~
     \nonumber\\= -p_{d} + \Lambda~~~~~~~~~~~~~~~~~~~~~~~~~~~~~
    \end{eqnarray}
   \begin{eqnarray}
 G^{0}_{0}=  3\frac{\dot{a^{2}}}{a^{2}}  + \frac{d(d-1)}{2}\frac{\dot{b^{2}}}{b^{2}}
  + 3d\frac{\dot{a}\dot{b}}{ab} ~~~~~~~~~~~~~~~~~~~~~~~~~~~~~~~~~~~~~~~~~~~~~~~~~~~~~~~\nonumber\\
  -\frac{1}{a^{2}}\left[2\frac{a''}{a} -
   \frac{a'^{2}}{a^{2}} + 4 \frac{a'}{ar} + d \frac{a'b'}{ab}
    + 2d \frac{b'}{br}
    + \frac{d(d-1)}{2}\frac{b'^{2}}{b^{2}}   + d\frac{b''}{b} \right]=
    \rho + \Lambda
   \end{eqnarray}
   where the (3+d+1)-dimensional stress-tensor, $T_{ij}$, in
   comoving co-ordinate should be of the form
\begin{equation}
T^{0}_{0}=\rho + \Lambda, T^{1}_{1} = T^{2}_{2} = T^{3}_{3} = -p +
\Lambda, T^{d}_{d} = -p_{d} + \Lambda
\end{equation}
 We have here assumed that $8\pi \bar{G} = 1 $ where $ \bar{G}$ is
 the gravitational constant in the higher dimensional space and a
 dot and and a prime overhead denote differentiation w.r.t. time
 and radial co-ordinate.

 To make the field equations tractable we assume, at this stage,
 that the scale factors $ a
 \equiv a(t) $ and $ b \equiv b(r,t) $
 and d = 1, i.e., we take a 5D spacetime where the $y$ = constant
  hypersurface is homogeneous FRW-like. In what follows we shall
 see that the fact that inhomogeneity is being introduced through
 the extra space has far reaching implications in the cosmological
 evolution of our models. For simplicity  as also the fact that
 the present universe is matter dominated, we consider here the
  \emph{dust} case only ($p = p_{d} = 0$).

 Now the equation (2) yields

\begin{equation}
b = a\beta (r) + \alpha (t)
\end{equation}

where $ \beta (r) $ and $ \alpha (t) $ are integration functions
of r and t respectively.
 Using equations (3), (4) and (8) we get
\begin{equation}
b = - c r^{2} a + \alpha (t)
\end{equation}
where c is a pure constant. However we subsequently see that c is
not exactly arbitrary, being a measure of the curvature of the 4D
space. Moreover $c = 0$ implies at once a flat and homogeneous
spacetime.
 In line with the existing literature
one assumes the dynamical behaviour of $\Lambda$ as $ \Lambda \sim
(\frac{\dot{a}}{a})^2, \Lambda  \sim \frac{\ddot{a}}{a}$ or $
\Lambda \sim \rho $. Although not strictly independent the first
one was proposed from dimensional consideration by Lima and Waga
\cite{lima} and subsequently adopted by several workers
\cite{ara}, \cite{pd}
 whereas the second type by Overduin and
Cooperstock \cite{ov} while  Vishwakarma \cite{vk} favoured  the
third alternative.

In our work we take $\Lambda =
kH^{2}$,$(H\equiv\frac{\dot{a}}{a})$ which through equation (5)
yields $a = t^{\frac{3}{(6-k)}}$ and $\Lambda =
\frac{9k}{(6-k)^{2}}t^{-2}$ . Now using equation (3) we get,
\begin{equation}
\ddot{\alpha} + \frac{6}{(6-k)}\dot{\alpha} t^{-1} -
3\frac{(3+k)}{(6-k)^{2}} t^{-2} \alpha
 + 4ct^{-\frac{3}{(6-k)}} = 0
 \end{equation}
 which through a long but somewhat straight forward calculation
 yields
\begin{equation}
\alpha = a_{1}t^{\frac{3}{(6-k)}}+a_{2}t^{-\frac{(3+k)}{(6-k)}}-
   \frac{2c(6-k)^{2}}{(12-k)(3-k)}t^{\frac{(9-2k)}{(6-k)}} \\
\end{equation}
$a_{1}$, $a_{2}$ are arbitrary constants.
 So clubbing everything together we finally get
 \begin{eqnarray}
   a &=& t^{\frac{3}{(6-k)}} \\
   b &=& ( a_{1}- cr^{2})t^{\frac{3}{(6-k)}}+a_{2}t^{-\frac{(3+k)}{(6-k)}}-
   \frac{2c(6-k)^{2}}{(12-k)(3-k)}t^{\frac{(9-2k)}{(6-k)}} \\
   \rho &=& \frac{\frac{9}{(6-k)}(a_{1} -
   cr^{2})t^{2\frac{(k-3)}{(6-k)}} +
   \frac{9(3-2k)}{(6-k)^{2}}a_{2}t^{\frac{12-k}{k-6}}-
   \frac{6ck(6-k)}{(12-k)(3-k)}}{(a_{1}-cr^{2})t^{\frac{6}{(6-k)}}
   + a_{2}t^{\frac{-k}{6-k}}-
   2c\frac{(6-k)^{2}}{((12-k)(3-k)}t^{2}} \\
   \Lambda &=& \frac{9k}{(6-k)^{2}}t^{-2}
 \end{eqnarray}

 One can now calculate the 4-space curvature $\left(R^{*(4)}\right)$ of the t-constant
 slice for the line element (1)$(d=1)$ through the
 expression, ~$R^{i}_{i} = R^{*(4)} + \dot{\theta} + \dot{\theta}^{2} -
2\omega^{2} + u^{i}_{;i}$ ~\cite{sc}, where $\theta$ is the
expansion scalar and last two terms refer to vorticity and
acceleration. After some algebra  we get the expression
\begin{equation}
R^{*(4)} = \frac{12c}{(a_{1} - cr^{2})t^{\frac{6}{(6-k)}} +
a_{2}t^{-\frac{k}{(6-k)}} - \frac{2ct^2(6-k)^{2}}{(12-k)(3-k)}}
\end{equation}

So as commented earlier, the arbitrary constant $c$ comes out to
be a measure of the curvature of the 4D space. $c$ is also a
measure of inhomogeneity parameter because $c = 0$ makes our
spacetime homogeneous. Relevant to mention that unlike the
analogous homogeneous case here the curvature also depends on
spatial co-ordinate $r$. Moreover it blows off at the initial
singularity $t=0$. Otherwise it is regular everywhere including
the origin of the distribution.

To check if our space time contains any geometric singularity
aside from the well known $t = 0$ big bang epoch we calculate the
Kretschmann scalar for our  5D inhomogeneous line element as
\begin{equation}
R^{ijkl}R_{ijkl} = 3 \frac{\ddot{a}^{2}}{a^{2}}
+\frac{\ddot{b}^{2}}{b^{2}} + 3\frac{\dot{a}^{4}}{{a}^{4}} +
2\left(\frac{\dot{a}}{a}\frac{\dot{b}}{b} -
\frac{1}{a^{2}}\frac{b'}{br} \right)^{2}
\end{equation}
At $t \rightarrow 0$, the invariant diverges but it is regular at
$r = 0$ since $\frac{b '}{r}$ is regular there. So unlike similar
cases in many inhomogeneous distributions there is no spatial
singularity in our cosmological model including the centre of
distribution. Moreover it has not escaped our notice that while
the 3D scale factor $a$ starts from $t=0$ the 5D scale $b$ is
infinite there. To ensure that both the scales start
simultaneously at the initial singularity we set the arbitrary
constant $a_{2}$ to be zero henceforth.\\

 The inhomogeneous
metric we presented here is very general in nature in the sense
that many well known solutions in this field
are recoverable as special cases of our metric.\\

(i) When $k = 0$ the $\Lambda$ term vanishes and we get back our
earlier solution \cite{dp}.

(ii)If $k = 0$ , $c = a_{1} = 0$, we get $a \sim t^{\frac{1}{2}}$,
$b \sim t^{-\frac{1}{2}}$ and $\rho$ = 0 which is the well known
vacuum solution of Chodos and Detweiler \cite{cd}.

(iii) Further for $c = a_{2} = 0$, we recover the isotropic
solution of Gr$\o$n  \cite{gr}.\\

If we now calculate the 3D deceleration parameter,
\begin{equation}
q = - \frac{\ddot{a}/a}{\dot{a}^{2}/a^{2}} = \frac{( 3-k )}{3}
\end{equation}
such that $ k > 3 $ for accelerating model. From equation (12) it
is evident that as we are discussing an expanding universe $k <6$.
Moreover, the second metric co-efficient b is regular when $k<
\frac{9}{2}$. Further, when $c > 0 $ and $ 0 < k < 3 $ we get
dimensional reduction but no acceleration. It also follows from
equation (13) that for $(a_{1} - cr^{2}) > 0$, $c > 0$ and $0 < k
< 3$, the mass density also remains non-negative as is also
evident from figure 1. Let us consider the situation for $c<0$. In
this case $\rho$ is always positive. If we further subdivide the
situation for (i) $0 < k < 3$ and (ii) $3 < k < \frac{9}{2}$, we
see for case (i) $\rho > 0$, but no dimensional reduction is
possible and no acceleration, for the case (ii) while $\rho$ is
again positive, we interestingly get both dimensional reduction
and acceleration. However small extra dimensions have of late been
somewhat out of favour following the resurgence of interests in
different brane inspired cosmological models where the extra
dimensions need not be small to account for the vexed hierarchial
problem of particle field theory.

It is conjectured that during the process of dimensional reduction
the extra dimensions finally stabilize at a very small length and
loose their dynamical character before the extra scale factor $b$
vanishes. Thereafter the cosmology enters the 4D phase without
having any reference to the extra dimensions \cite{sahdeb}. So the
second singularity at $b=0$ is never reached and the invariants of
equations (16-17) as also the mass density never have the chance
to diverge there. For this model this transition has far reaching
implications because the very existence of the extra space, so to
speak, seems to induce inhomogeneity in this case. So not only do
we enter a 4D era, it also envisages a smooth transition  from a
multidimensional, inhomogeneous phase to a 4D homogeneous one.
Interestingly this transition takes place without forcing
ourselves to choose very special initial conditions as is the
practice in the four dimensional cosmology. This, in our opinion
is a very important aspect of our model. So the primordial
inhomogeneities die down in
a natural way.\\
\begin{figure}[h]

\begin{center}

 \includegraphics[width=6 cm]{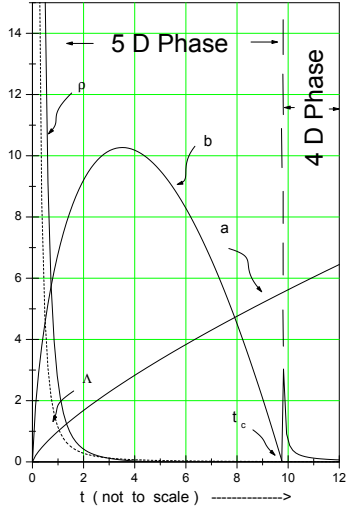} \caption{
 \small \emph{The time evolution of a, b, $\rho$ and $\Lambda$ are
 shown in this figure for $c>0$ and $0<k<3$ (say, $(a_{1} -cr^2) = 10,
 ~k = 2,~a_{2} = 0$  and $c = 1$ ). In this case dimensional
  reduction is possible but no acceleration. At
  compactification $t = t_{c}$, $\rho$ surges to infinity and cosmology enters 4D phase}\label{1}
    }
 \end{center}
\end{figure}

\section*{3. Homogeneous and Isotropic Universe}

~~~~~~As our universe is now manifestly homogeneous and isotropic
we assume a = b and also the metric coefficients depend on time
alone. These assumptions also facilitate to make comparison with
the observational results as also to gauge the influence of extra
dimensions on the recent observational findings.

With these two assumptions our general field equations (1) now
reduce to
\begin{eqnarray}
  \frac{(d+2)(d+3)}{2}\frac{\dot{a}^{2}}{a^{2}} &=& \rho + \Lambda \\
  (d+2)\frac{\ddot{a}}{a} + \frac{(d+1)(d+2)}{2}\frac{\dot{a}^{2}}{a^{2}}&=& \Lambda
\end{eqnarray}
As there are two independent equations with three unknowns we once
again assume $ \Lambda = kH^{2} $. With this assumption we finally
get,
\begin{eqnarray}
  a &=& ( Ct + D )^{\frac{2(d+2)}{(d+2)(d+3)-2k}} \\
  \rho &=& \frac{2(d+2)^{2}}{(d+2)(d+3)-2k}t^{-2}\\
  \Lambda &=& \frac{4k(d+2)^{2}}{\left[(d+2)(d+3)-2k \right]^{2}}t^{-2}\\
  H &=& \frac{2(d+2)}{(d+2)(d+3)-2k}t^{-1}
\end{eqnarray}
we can set the constant $C = 1$ and $D = 0$ without loss of
generality in equation (21). Further the positivity of $ \rho$
gives the restriction $ k< \frac{(d+2)(d+3)}{2}$

As a special case to our solution (21) we see that

(i) $k = 0$ ( i.e. , $\Lambda = 0 $ ) implies $ a =
t^{\frac{2}{d+3}} $ which is the higher dimensional generalization
of the well known Einstein- de Sitter solution.

(ii) further for the usual 4D case ($ d = 0$) $ a \sim
t^{\frac{2}{3}}$ which is the FRW matter dominated case.

(iii) on the other hand $d = 0$, but $ \Lambda \neq 0 $ , $ a \sim
t^{\frac{2}{3 - k}}$ \cite{sr}\\

\textbf{Characteristics of the Model}\\

From equation (19)
\begin{equation}
\frac{\rho}{\frac{(d+2)(d+3)}{2}H^{2}} +
\frac{\Lambda}{\frac{(d+2)(d+3)}{2}H^{2}} = 1
\end{equation}

or $\Omega_{m} + \Omega_{\Lambda} = 1$ ( in the absence of
curvature term ) where $\Omega_{m}$ is the higher dimensional
cosmic matter density parameter and $\Omega_{\Lambda}$ is the
corresponding vacuum energy density parameter. From equations (21)
- (25) it also follows that $ \Omega_{m} =
\frac{(d+2)(d+3)-2k}{(d+2)(d+3)} $ and $ \Omega_{\Lambda} =
\frac{2k}{(d+2)(d+3)}$ such that $\Omega_{m} + \Omega_{\Lambda} =
1$. So our solutions are consistent. It also follows from the
expression of $ \Omega_{\Lambda} $ that the arbitrary constant k
is also a measure of the cosmic vacuum density parameter. We
further see that the deceleration parameter comes out to be

\begin{equation}
q = - \frac{\ddot{a}}{aH^{2}}= \frac{(d+1)(d+2)-2k}{2(d+2)} =
\frac{(d+1) - (d+3)\Omega_{\Lambda}}{2}
\end{equation}

For our cosmology to accelerate $ k > \frac{(d+1)(d+2)}{2}$ such
that we finally restrict $k$ as $ \frac{(d+1)(d+2)}{2} < k <
\frac{(d+2)(d+3)}{2} $. This inequality ensures that we get both
positivity of $ \rho$ and acceleration. To make our analysis
consistent with the present day observational results we find that
for $q = -0.5$, $ k = \frac{(d+2)^{2}}{2} $. For the particular
case of 5D universe ($d= 1$), taking the most acceptable value of
$ H_{0}$ = 72 km/s/mpc we calculate the age of the universe to be
equal to 25.5 Gyr, quite large. Almost same age was obtained by
Viswakarma \cite{vk1} in a different context. Another point to
mention is that the above value of k implies  (for $d  = 1$)  $
\Omega_{m0} = 0.25 $ and $\Omega_{\Lambda 0} = 0.75$, which are
fairly consistent with current observational results.

Another attractive feature of our model is that it gives a
sufficiently large age of the Universe as mentioned earlier. It is
evident from our equation (24) that the age of the universe for
our model comes out to be  $ t_{0} = \frac{3}{(6 - k)H_{0}}$ (for
$d = 1$). This is very remarkable in view of the fact that the age
of the universe in the FRW model with a constant $\Lambda$ is
uncomfortably close to the age of the globular clusters $t_{GC} =
12.5 \pm 1.2 $ Gyr \cite{cr}. The quintessential models give even
lower age. As there is considerable uncertainty in the exact value
of $H_{0}$ we have calculated $\Omega_{\Lambda 0}$ and $ t_{0}$
for different value of k, keeping $H_{0}$ a constant for a
particular table as shown below.
\begin{table}[h]
\caption{\emph{Age of the universe for different values of $H_{0}$
(for d = 1)}  }
\begin{center}

   \begin{tabular}{ c c c c c c c c}\hline\hline

k&q&$\Omega_{\Lambda 0}$&$\Omega_{m 0}$&t(Gyr)&t(Gyr)&t(Gyr)\\
&&&&$H_{0}$ = 72 km/s/mpc&$H_{0}$ = 80 km/s/mpc&$H_{0}$ = 100
km/s/mpc\\\hline
0.00& 1.00&0.00&1.00&6.25&5.63&4.50\\\
1.00& 0.67&0.17&0.83&7.50&6.75&5.40\\\
2.00& 0.33&0.33&0.67&9.38&8.44&6.75\\\
3.00& 0.00&0.50&0.50&12.50&11.25&9.00\\\
3.50&-0.17&6.75&8.44&11.25&13.50&16.88\\\
4.00 &-0.33&0.67&0.33&18.75&16.88&13.50\\\
4.50&-0.50&0.75&0.25&25.00&22.50&18.00 \\\hline\hline
\end{tabular}
\end{center}
\end{table}

Next we have plotted the age of the universe $t_{0}$ against
$\Omega_{\Lambda 0} $ on the basis of our model. It is found that
as $\Omega_{\Lambda 0} $ increases $t_{0}$ also increases. If the
required mass density $\Omega_{m 0} $ was smaller or
$\Omega_{\Lambda 0} $  is larger one could get higher age in these
model as is clear from figure 2.
\begin{figure}[h]
\begin{center}
  \includegraphics[width=6cm]{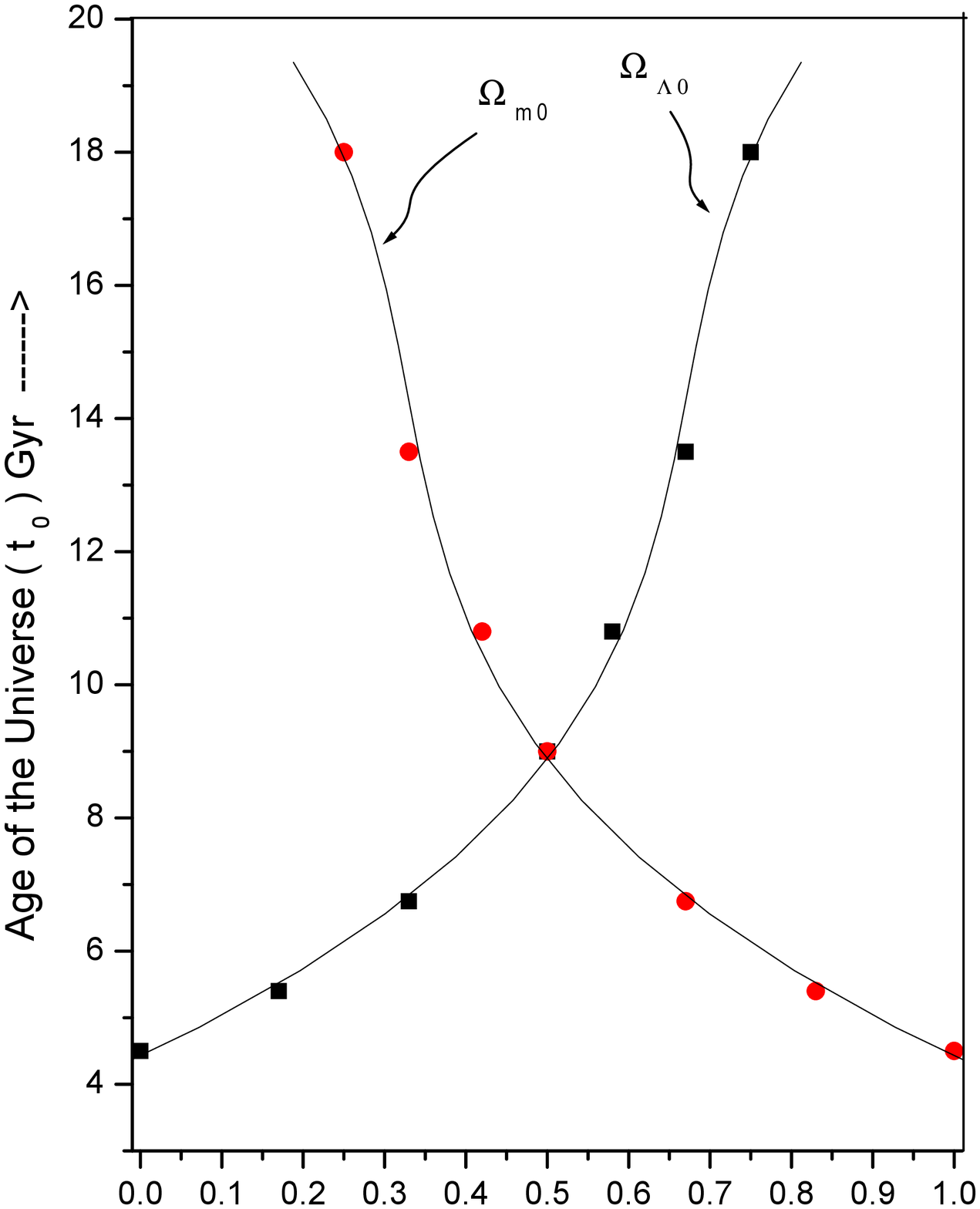}
  \caption{
 \small \emph{ As $\Omega_{\Lambda 0}$ increases
 ( or $\Omega_{m 0}$ decreases ) $t_{0}$  increases }\label{1}
    }
 \end{center}
\end{figure}

Again we see from the table 1 that the gradual increases of $H_{0}
$ provide the more realistic age of the Universe, for $H_{0}$ =
100 km/s/mpc, the age of the universe $t_{0}$ = 18 Gyr whereas the
most acceptable age of being 14 Gyr \cite{sr}. We also see from
the  table 1 that for k = 3, q = 0, i.e. , the flip occurs for
this value of k. For $H_{0} $ = 100 km/s/mpc, the flip time $t_{f}
$ = 9 Gyr, i.e., before it the universe decelerates. So our model
$ \Lambda \sim H^{2}$ is amenable to past deceleration and current
acceleration.

\section*{4. Influence of Dimension  }

\textbf{(i) Age of the Universe : }

If we calculate $ t_{0} = \frac{2}{(d+3)\Omega_{m0}H_{0}} $ (see
equation (24)), taking the most acceptable values of $\Omega_{m0}$
= 0.33 and $H_{0} = 72\pm 8~ km/s/mpc$ and then plot $t_{0} \sim d
$ curve the age of  the universe is found to decrease with the
number of dimensions of the universe. This is a very interesting
result of our whole analysis ( see figure 3).
\begin{figure}[h]
\begin{center}
  \includegraphics[width=5cm]{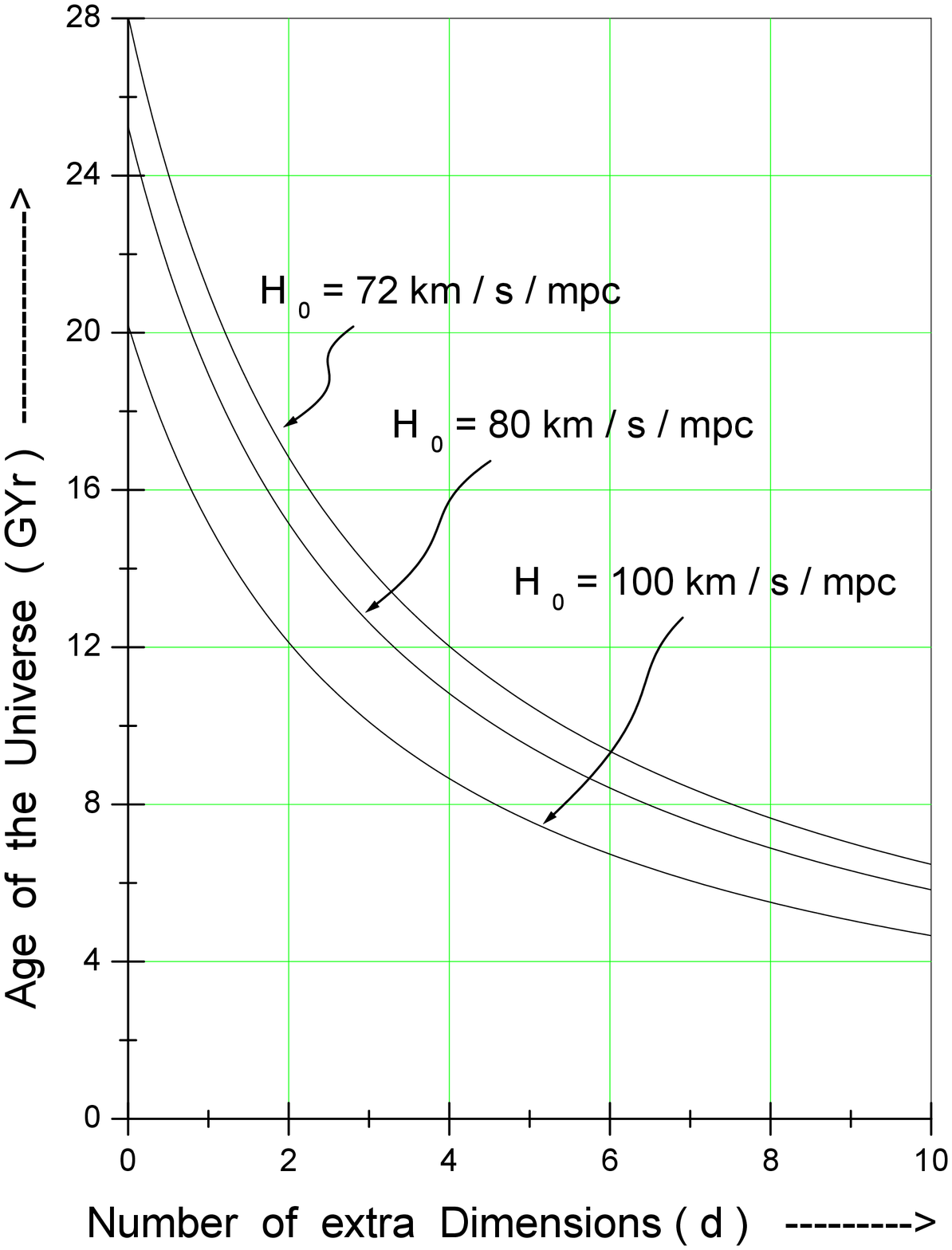}
  \caption{
 \small \emph{This figure shows that as the number of dimension
 increases age of the universe decreases}\label{1}
    }
 \end{center}
\end{figure}

\textbf{(ii) Cosmological constant :}

~~~~~The relation between $\Lambda$ with the number of  dimensions
of the universe is written as
\begin{equation}
     \Lambda = \frac{1}{2}H_{0}^{2}( 1 - \Omega_{m0})(d+2)(d+3)
\end{equation}
 If we plot $\Lambda \sim d $ one sees that $\Lambda$ increases
  as the number of extra dimensions (d ) increases (figure 4).
 \begin{figure}[h]
\begin{center}
  \includegraphics[width=5cm]{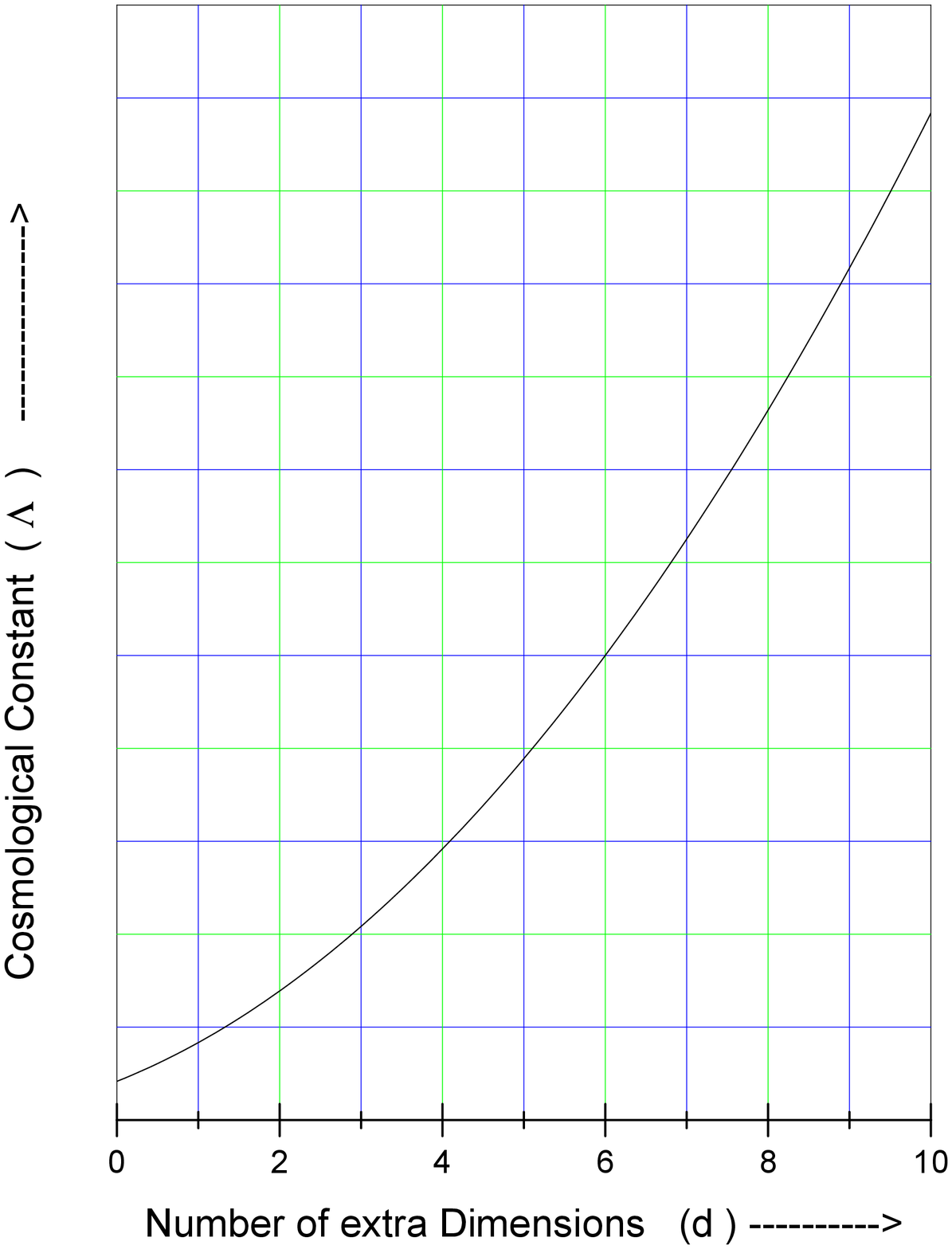}
  \caption{
 \small \emph{As the number of dimension
 increases $\Lambda$ also increases }\label{1}
    }
 \end{center}
\end{figure}

\textbf{(iii) Deceleration Parameter :}

\begin{equation}
     q = \frac{1}{2}\left[(\Omega_{m0} d - 3(1 - \Omega_{m0})\right]
\end{equation}

The above equation implies that as the number of dimensions
increases $q$ increases.  For constant $ \Omega_{m0}$ it is shown
that  $q \propto d $. The  figure 5 shows that (for $\Omega_{m0}$
= 0.33 ) as the number of dimensions increases, $q$ increases.
Obviously the flip is delayed with the increase of number of
dimensions for any particular value of $\Omega_{m0}$.
\begin{figure}[h]
\begin{center}
  \includegraphics[width=5cm]{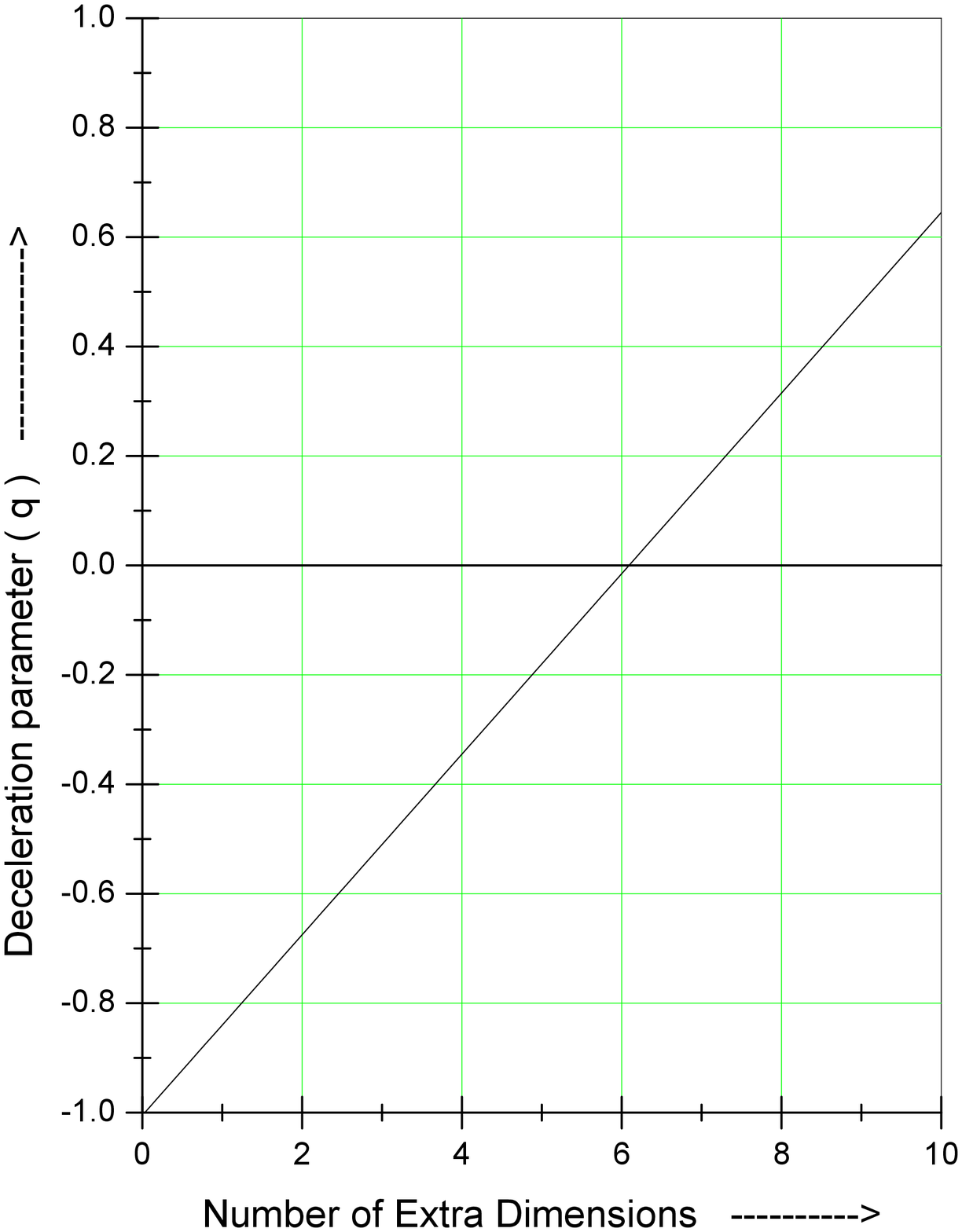}
  \caption{
 \small \emph{q increases with the number of dimensions
    }\label{1}}
 \end{center}
\end{figure}
\section*{5. Some Astrophysical Parameters in our model : }

~~~~~~In this section we shall very briefly discuss the proper
distance D(z), luminosity distance ($D_{L}$) and look back time.
As these quantities are extensively discussed for 4D space time in
the standard text books \cite{nar} we shall omit detailed
mathematical calculations and give only the final results
generalised to ( 4 + d ) dimensions for our dust model. \\

\textbf{(i) Proper Distance :} Let a star located at  $r=r_{0}$
emit radiation at  $t=t_{1}$ and an observer at $r = 0$ receive
the same at  $t=t_{0}$. The path being evidently a null geodesic
($\theta_{1},\theta_{2},\theta_{3} .....,\theta_{n} = constants$).
The  proper distance between the source and observer is given by
\begin{equation}
    D(z) = a_{0}\int_{a}^{a_{0}}\frac{da}{a\dot{a}} =
    H_{0}^{-1}\left[\frac{2(d+2)}{(d+2)(d+1)-2k}\right]\left[1 - (1+z)^{\frac{2k-(d+2)(d+1)}{2(d+2)}}\right]
\end{equation}
for small z it reduces to,
\begin{equation}
    H_{0}D(z) = z-\frac{(d+2)(d+3)-2k}{4(d+2)}z^{2} + .....
    = z - \frac{1}{2}( 1 + q ) z^{2} + ...
\end{equation}

Again, for $z \rightarrow \infty$ but no acceleration i.e., $k<\frac{1}{2}(d+1)(d+2)$,\\
$D(z=\infty) = H_{0}^{-1}\frac{2(d+2)}{(d+1)(d+2)-2k}=
\frac{H_{0}^{-1}}{q} = \frac{H_{0}^{-1}}{2\Omega_{m0}-1} $,
for acceleration obviously $ D(z) \rightarrow\infty$ \\

\textbf{(ii) Luminosity Distance $D_{L}$ :} Luminosity distance is
another very useful concept of relativistic astrophysics. If
$D_{L}$ is the luminosity distance of the object, L being the
total energy emitted by a Galaxy in unit time then
\begin{equation}
    D_{L} = \left(\frac{L}{4\pi l}\right)^{\frac{1}{2}}
\end{equation}
After some straight forward calculation we get for our model
\begin{equation}
H_{0} D_{L} =
\frac{2(d+2)}{(d+1)(d+2)-2k}(1+z)\left[1-(1+z)^{\frac{2k-(d+1)(d+2)}{2(d+2)}}\right]
\end{equation}

for small z,
\begin{equation}
H_{0} D_{L} = z+\frac{1}{2}(1-q)z^{2} = z + \frac{1}{4}\left[5 -
\Omega_{m0}(d+1)\right]z^{2}
\end{equation}
\begin{figure}[h]
\begin{center}
  \includegraphics[width=5cm]{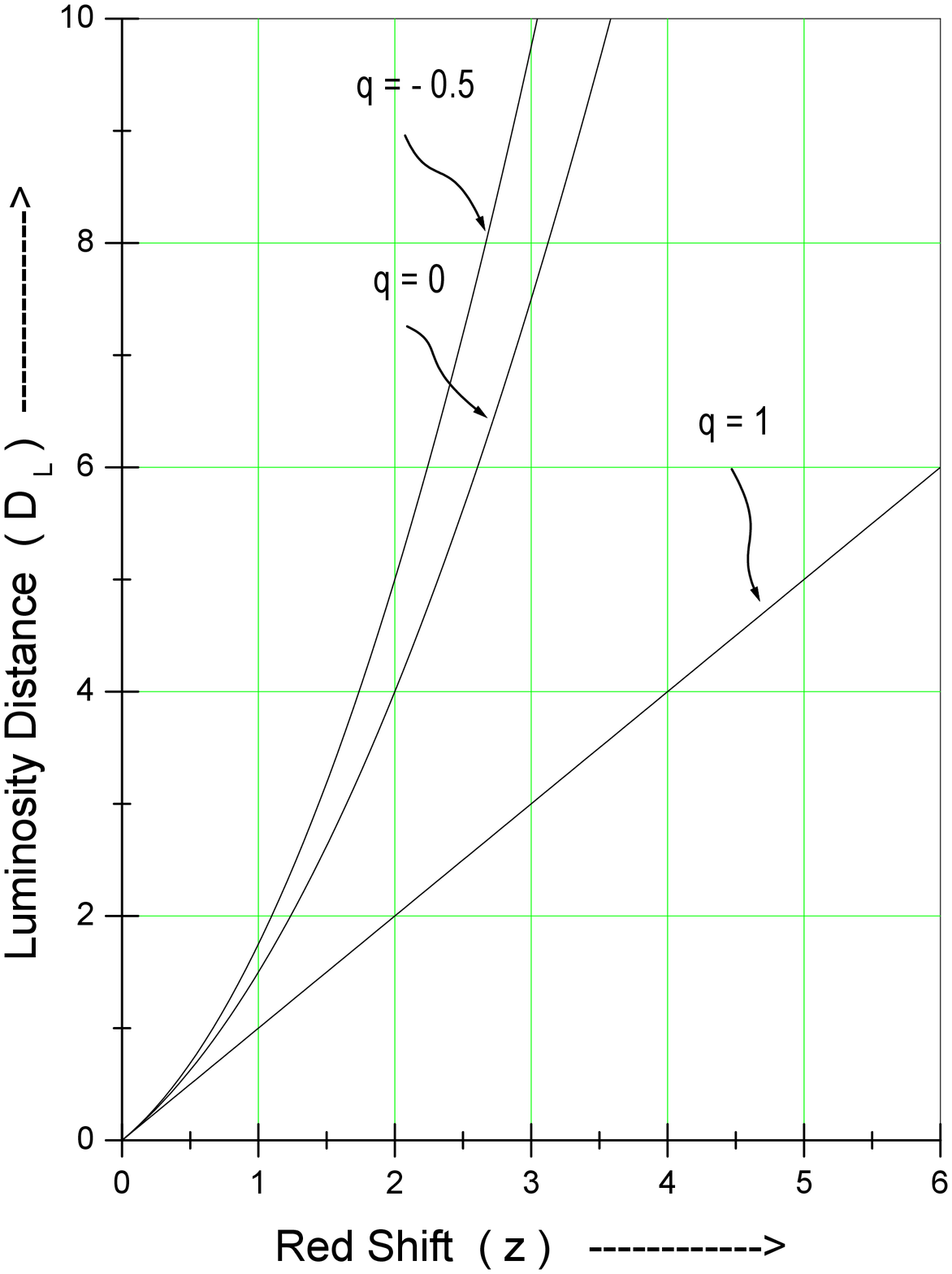}
  \caption{
 \small \emph{The Luminosity distance $D_{L}$ expressed as a function of the
 redshift $z$ for $q = 1, 0, -0.5$. The relationship is linear, as predicted
 by Hubble's linear Law. For $q_{0} = - 0.5$, $D_{L}$ increases  faster.
  All curves merge for small z }\label{1}
    }
 \end{center}
\end{figure}
We have plotted  $D_{L} ( q_{0}, z )$ as a function of $z$ for
various parametric values of $q_{0}$ in the figure 6. Note that
all curves start off with the linear Hubble law $z = H_{0}D_{1}$
for small value of $z$, but then fan out, with only the curve for
$q_{0} = 1$ staying linear all the way. As a rule we notice that,
for the same redshift, the luminosity distance is larger for lower
values of $q_{0}$. Thus for $q_{0} = 1$, we have, $D_{L} =
\frac{z}{H_{0}}$ whereas for $q_{0} = 0$, we get $D_{L} =
\frac{z}{H_{0}}[1 + \frac{1}{2}z]$, again for $q_{0} = -0.5, D_{L}
= \frac{z}{H_{0}}\left(1+\frac{3}{4}z\right)$
 \\
 \textbf{(iii) Look back Time :}
 The time in the past at which the light we now receive
from a distant object was emitted is called the look back time.
The radiation travel time ( or look back time ) $(t - t_{0})$
generalised to higher dimensions for photon emitted by a source at
instant t and received at $t_{0}$ is given by,
\begin{equation}
t-t_{0} = \int_{a}^{a_{0}}\frac{da}{\dot{a}}
\end{equation}
In our case, $a = t^{\frac{2(d+2)}{(d+2)(d+3)-2k}}$
so,
\begin{equation}
H_{0}(t_{0} - t) =
\frac{2(d+2)}{(d+2)(d+3)-2k}\left[1-(1+z)^{\frac{(d+2)(d+3)-2k}{2(d+2)}}\right]
\end{equation}
For small z the above equation reduces to,
\begin{eqnarray}
H_{0}(t_{0} - t)&=&  \left[z - \frac{(d+2)(d+3)-2k}{2(d+2)}z^{2} +
..
.   \right]\nonumber\\
& = & z\left[1-\frac{1}{2}(d+3)\Omega_{m}z+ ... \right]
\end{eqnarray}
\begin{figure}[h]
\begin{center}
  \includegraphics[width=5cm]{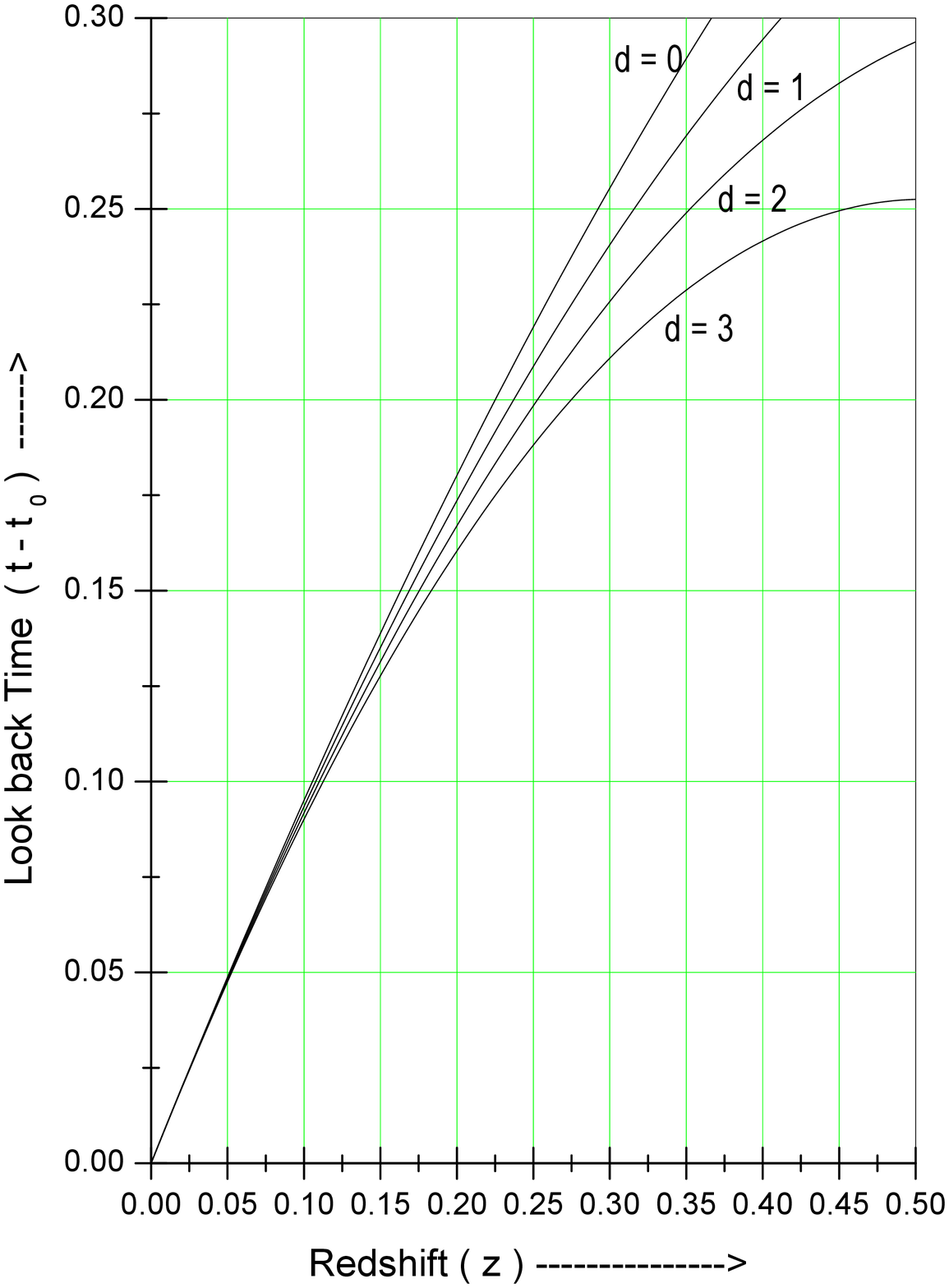}
  \caption{
 \small\emph{Look back time increases with red shift (for small value of z ) }\label{1}  }
\end{center}
\end{figure}
It follows from above that for a fixed z the look back time
decreases with number of dimensions as evident from figure 7.
Again it is shown that the look back time also increases with red
shift (for small value of z).
\\
\section*{6. Cosmological model with constant $\Lambda $ : }

~~~~~~For the sake of completeness as also of relevance to the
current accelerating scenario we briefly discuss in this section
an earlier work of one of us with a constant $\Lambda$
\cite{scbb}. In this work we discussed first zero pressure
inhomogeneous 5D model with a constant $\Lambda$ and later
generalized it to a fluid obeying an equation of state $p = m\rho
\neq p_{5}$ where p is the isotropic three pressure, $ p _ {5}$ is
the pressure in the fifth dimension, $\rho$ is the matter density
and $m$ is a constant. At this stage it will not be out of place
to point out that our set of solutions in Section 2 does not
automatically reduce to the above mentioned works with a constant
$\Lambda$ as our  ansatz $\Lambda = kH^{2}$ in the present work
does not permit that type of reduction.

~~~~~~For the line element (1) with d=1 ( 5D space time) and
$\Lambda$ = constant we get for the matter dominated case ($p =
p_{d} = 0$)
\begin{equation}
a = a_{0}\sinh^{\frac{1}{2}}pt
\end{equation}
\begin{equation}
b = \frac{(b_{0}-cr^{2})\sinh pt + b_{1}\cosh pt +
\frac{4c}{p^{2}}}{\sinh^{\frac{1}{2}}pt}
\end{equation}
 $p=\sqrt{\frac{2\Lambda}{3}}$, where ($a_{0}, b_{0}, b_{1}$ are
 arbitrary constants, and c is the usual 4D curvature in the
 t-constant hypersurface.

 To facilitate comparison with our homogeneous space of section
 (3) we put c = 0 such that the line element read
\begin{eqnarray}
a = a_{0}\sinh^{\frac{1}{2}}pt  ;~~~~
 b  = \frac{b_{0}\sinh pt +
b_{1}\cosh pt }{\sinh^{\frac{1}{2}}pt}
\end{eqnarray}
Following \cite{rom} we, at this stage, define a deceleration
parameter such that an affective `scale factor' L becomes
\begin{eqnarray}
L = (a^{3}b)^{\frac{1}{4}} = \sinh ^{\frac{1}{4}}pt ( \sinh pt +
\cosh pt )^{\frac{1}{4}}
\end{eqnarray}
( putting $a_{0},b_{0},b_{1}$ unity, for simplicity). The well
known 4D deceleration parameter ( say $q_{1}$ ) in this formalism
becomes
\begin{eqnarray}
q_{1} = - \frac{\ddot{L}L}{\dot{L}^{2}} = -8 \sinh pt \cosh pt + 8
\cosh^{2} pt - 5
\end{eqnarray}
The behaviour of $q_{1}$ here is interesting. At the early epoch
the last expression shows that $q_{1} > 0$ ( good news for
structure formations ).

The flip occurs at the instant $t = t_{0}$, then the universe
starts acceleration in conformity with the present day
observation. It should, however, be emphasized that any attempt to
compare the flip time of our model given above with observational
findings will be too ambitions due to several initial assumptions
in solving the equations as also due to the presence of so many
arbitrary constants. One may, however, fine tune the arbitrary
constants to account for the current observations.

If we further assume $b_{1} = 0$ we get isotropic expansion , i.e.
, $a = b$. In this case the expression for q ( say $q_{2}$) turns
out to be
\begin{eqnarray}
q_{2} = -\frac{\cosh^{2} pt-2}{\cosh^{2} pt}
\end{eqnarray}
From the expression (41) and (42) it is evident that the flip
occurs earlier for the anisotropic case, which is also evident
from our figure 8. However the cosmological implications of this
finding need further investigation.
\begin{figure}[h]
\begin{center}
  \includegraphics[width=5cm]{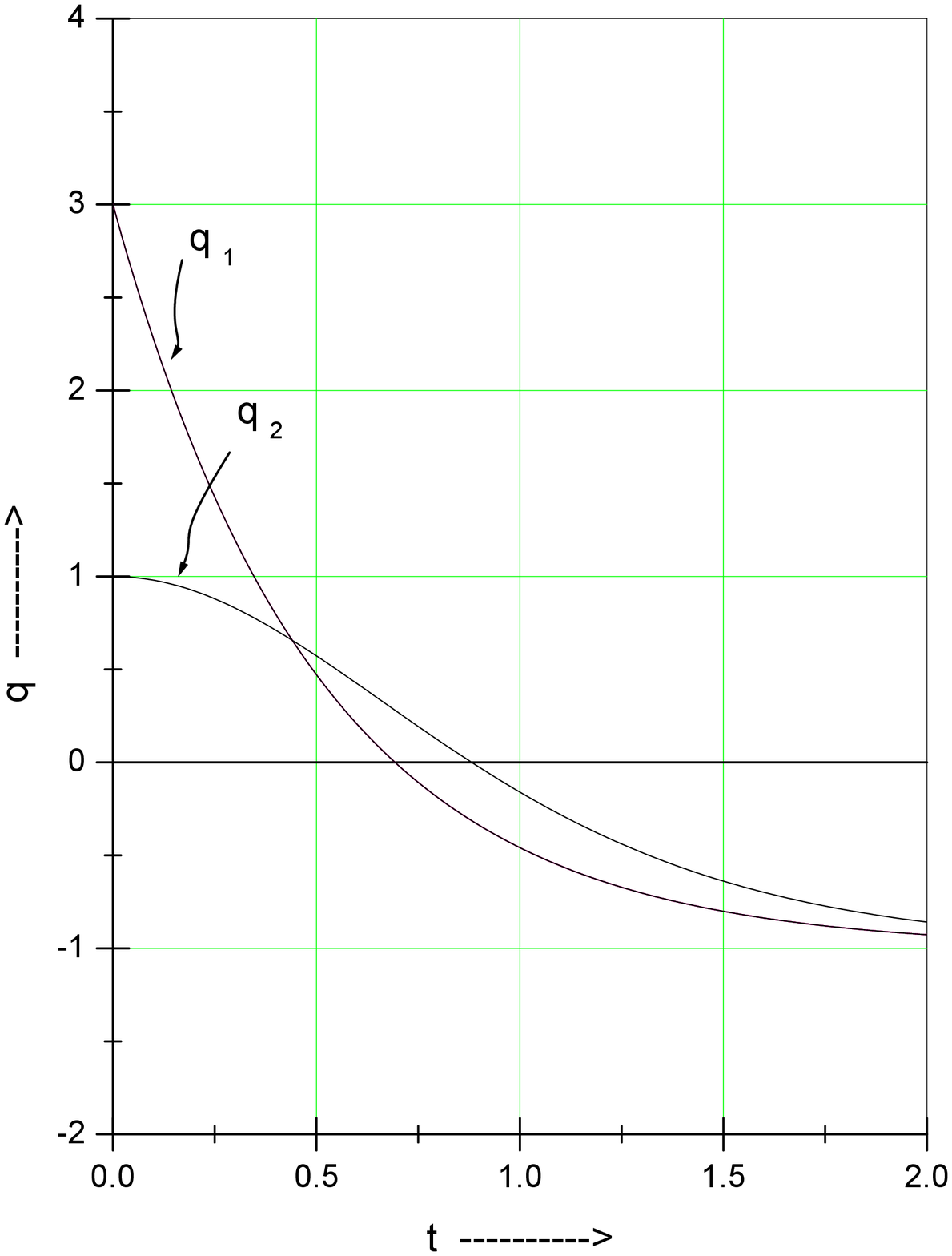}\\
  \caption{
 \small\emph{The time evolution (t not to scale) of q is shown in this figure.
It is shown that flip for $q_{1}$ is earlier than flip for
$q_{2}$.  }\label{1} }
\end{center}
\end{figure}

To end the section let us find out if the presence of extra
dimensions, in any way, influences the instant of flip for the
isotropic case. From the divergence equation in
(3+d+1)-dimensional homogeneous space time for line element (1),
we get,
\begin{eqnarray}
\dot{\rho} + \left(3\frac{\dot{a}}{a} +
d\frac{\dot{b}}{b}\right)\left( \rho + p\right) = 0
\end{eqnarray}
which for our case ($p = 0,~ a= b$) yields
\begin{eqnarray}
\rho = Ca^{-(d+3)}
\end{eqnarray}
Using equation (44) in equation (19), we get

\begin{eqnarray}
\frac{\dot{a}^{2}}{a^{2}} = \frac{2}{(d+2)(d+3)}\left[\Lambda +
Ca^{-(d+3)}\right]
\end{eqnarray}
 Skipping mathematical
details and adjusting the arbitrary constant $C$ we finally  get
from equation (45 )
\begin{eqnarray}
a = a_{0} \sinh^{\frac{2}{d+3}} pt
\end{eqnarray}
where $p = \sqrt{\frac{d+3}{d+2}\Lambda}$. So the deceleration
parameter reduces to
\begin{eqnarray}
q = \frac{(d+3)-2 \cosh^{2} pt}{2 \cosh^{2} pt}
\end{eqnarray}
so the flip occurs at
\begin{eqnarray}
\cosh pt_{c} = \sqrt{\frac{d+3}{2}}
\end{eqnarray}
Obviously the flip is delayed with the increasing of number of
dimensions  although a physical explanation of this behaviour is
beyond the scope of the present investigations. Interestingly we
got the same result in section 4 for the case of varying $\Lambda$
( figure 9).
\begin{figure}[h]
\begin{center}
  \includegraphics[width=5cm]{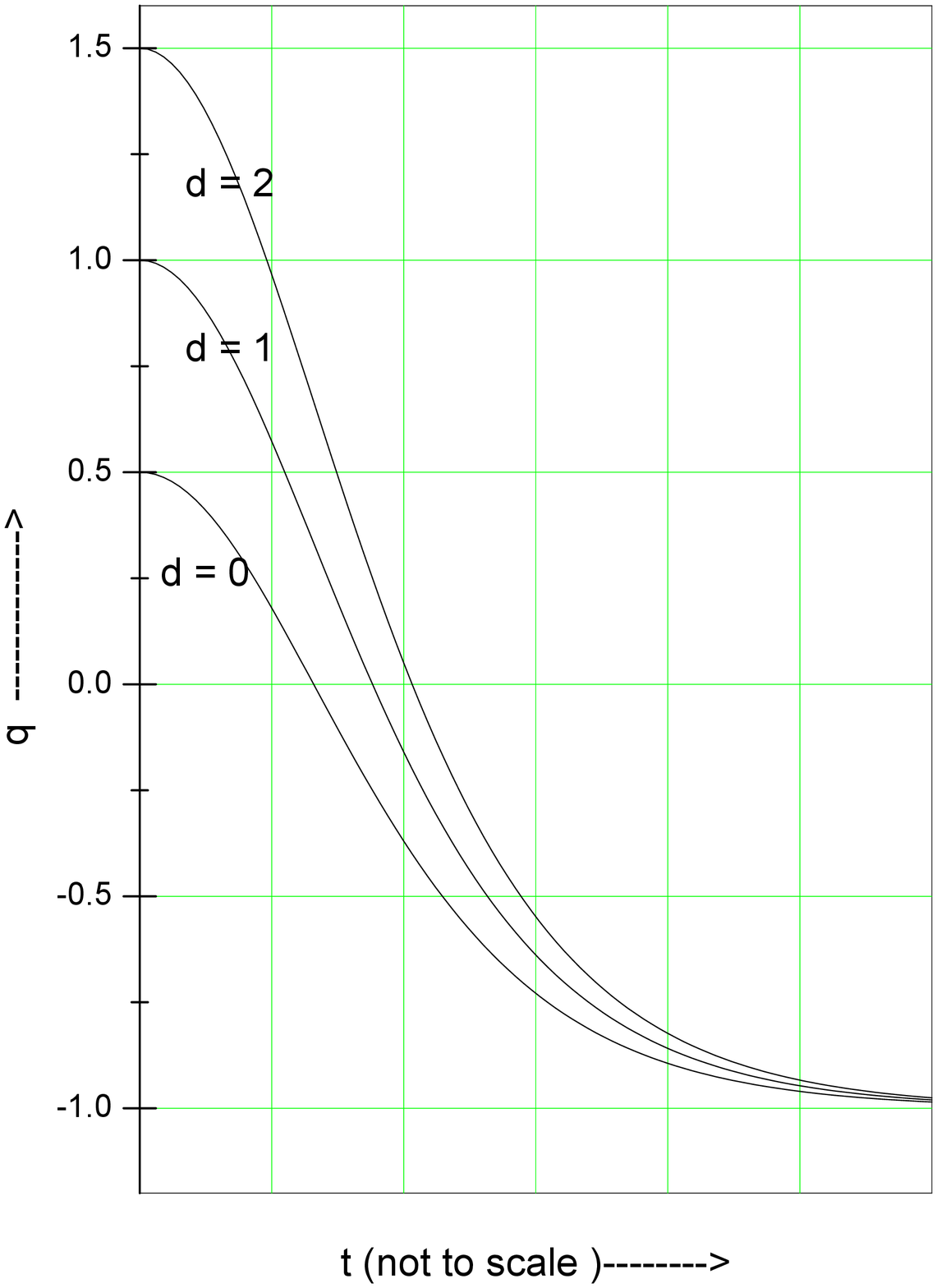}\\
  \caption{
 \small\emph{ Flip is delayed with the number of
dimensions }\label{1} }
\end{center}
\end{figure}

\section*{7. Discussion }

    ~~~~~~In this work we have studied a multidimensional model (both
homogeneous and inhomogeneous) with a time varying cosmological
parameter. It is found that under certain conditions the cosmology
gives an accelerated expansion. In the inhomogeneous case no
initial deceleration is found. However for the homogeneous case an
initially decelerating universe starts accelerating undergoing a
flip.This is good for both structure formation and current
observational status. Some of our cases are also amenable to the
desirable feature of dimensional reduction. As mentioned earlier
the dimensional reduction for our inhomogeneous model seemingly
points to the fact that a primordial inhomogeneous higher
dimensional cosmology enters a 4D homogeneous era although we do
not have to make any stringent initial conditions to achieve this
homogenization as tn the standard 4D case. This is an important
finding of our analysis.While it is conjectured that the extra
space stabilises at planckian length it is not apparent from our
analysis how that mechanism in the form of a sort of repulsive
potential field actually works in our model.This is definitely a
defect of our model.However in an earlier work Guendelmann and
Kaganovich\cite{gk} studied the Wheeler- Dewitt equation in the
presence of a negative cosmological constant and dust and showed
that quantum effects do stabilise the volume of the universe, thus
providing a mechanism of quantum avoidance of the singularity.
Other two important findings in our analysis are that the time of
flip depends on the number of extra dimensions. In fact as the
number of dimensions increase the flip is delayed thus higher
dimensional models allow more time for structure formation.
Secondly the flip also depends on anisotropy. Shear simply hastens
the arrival of the instant of flip. It is too premature to attempt
a physical explanation of the above two results but the results
are interesting enough to warrant further investigations in this
regard. To end a final remark is in order. We mentioned in the
introduction that both extra dimensions and inhomogeneities
produce a set of extra terms to the 4D FRW model, which creates a
sort of back reaction \cite{kari} in the process. While the effect
of extra dimension is discussed here in some detail, the influence
of inhomogeneities in higher dimensional model will be discussed
in our future work.\\
To end a final remark may be in order. We have here presented
results which include cases when the extra dimensions also inflate
along with the ordinary ones in section 3. While the idea of large
extra dimensions is not as repugnant these days as in the past
following its new found relevance in brane models but an inflating
extra dimension goes far beyond that.This is a serious shortcoming
of the present analysis, particularly the form of the spacetime we
chose in section 3. While  working in higher dimensional spacetime
one should see that the extra space should form a compact manifold
with the symmetry group G so that the $(d+3)$ spatial symmetry
group is a direct product $O(3)\times G$ and not $O(d+3)$,
although there exists a good number of works in this type of space
also.\\

\textbf{Acknowledgment : } We acknowledge the financial support of
UGC, New Delhi as also valuable comments of the anonymous referee.

\end{document}